\documentclass{article}

\usepackage{graphicx}

\newcommand{\nue}{\nu_e}

\title{Physics Reach of Electron-Capture Neutrino
    Beams\footnote{Presented at the 7th International Workshop on
    Neutrino Factories and Superbeams (NuFact 05), Frascati, Italy,
    June 2005. The article will appear in the Conference
    Proceedings.}}

\author{
J.~Bernabeu$^{\rm a}$, J.~Burguet-Castell$^{\rm a}$,
C.~Espinoza$^{\rm a}$, M.~Lindroos$^{\rm b}$ \\
\small{$^{\rm a}$Universitat de Valencia and IFIC, E-46100 Burjasot, Spain.} \\
\small{$^{\rm b}$AB-department, CERN, Geneva, Switzerland.}
}

\date{}

\begin{document}

\maketitle

\begin{abstract}
To complete the picture of neutrino oscillations two fundamental
parameters need to be measured, $\theta_{13}$ and $\delta$. The next
generation of long baseline neutrino oscillation experiments
--superbeams, betabeams and neutrino factories-- indeed take aim at
measuring them. Here we explore the physics reach of a new candidate:
an electron-capture neutrino beam. Emphasis is made on its feasibility
thanks to the recent discovery of nuclei that decay fast through
electron capture, and on the interplay with a betabeam (its closest
relative).
\vspace{1pc}
\end{abstract}

\section{Introduction}

Electron Capture is a process in which an atomic electron is captured
by a proton of the nucleus leading to a nuclear state of the same mass
number $A$, replacing the proton by a neutron, and emitting an
electron neutrino: $p ~ e^- \to n ~ \nue$.

It is a two body decay, and so the energy of the resulting neutrino is
fixed. Thus, if we accelerate ions unstable by electron capture and
let them decay in straight sections pointing to a far detector, we get
a \emph{pure} (that is, only one neutrino flavor) and
\emph{monochromatic} neutrino beam.

The rationale for doing this is much the same as for the $\beta$-beam,
but with a nice extra feature which is to have a very peaked energy
spectrum.

In reference \cite{Bernabeu:2005jh} there is a more thorough
discussion of the results we present here.

\section{Feasibility}

A possible implementation of the concept would involve the use of
EURISOL to produce the unstable ions, the SPS to accelerate them, and
a decay ring, much like the one proposed for $\beta$-beams
\cite{Autin:2002ms}. However, to allow electron capture to occur, we
need to keep one electron bounded to the ion, and partly charged ions
have a short vacuum life-time (even in a very good vacuum collisions
with the few remaining atoms suffice to make them quickly lose its
electron). So we need ions that decay \emph{fast enough}.

Recent discovery of nuclei far from the stability line, having super
enhanced spin-isospin transitions to a giant Gamow-Teller resonance
kinetically accessible \cite{Algora:2004vv} turn out to be very good
candidates. This discovery of nuclei that do decay fast enough through
electron capture is the cornerstone that opens a window for a
monochromatic neutrino beam experiment.

In Table \ref{table:nuclei} we show the properties of a few ion candidates.

\begin{table*}
  \caption{Decay properties of some rare-earth nuclei.}
  \label{table:nuclei}
  \begin{tabular}{|c|c|c|c|}
\hline
Decay                      & $T_{1/2}$ &  $E_\nu$ (keV) & EC/$\beta^+$ (\%) \\
\hline
$^{148}$Dy $\to$ $^{148}$Tb       &  3.1 m   &  2062     &   96/4      \\
$^{150}$Dy $\to$ $^{150}$Tb       &  7.2 m   &  1397     &   99.9/0.1  \\
$^{152}$Tm $2^-$ $\to$ $^{152}$Er &  8.0 s   &  4400     &   45/55     \\
$^{150}$Ho $2^-$ $\to$ $^{150}$Dy &   72 s   &  3000     &   77/33    \\
\hline
  \end{tabular}
\end{table*}

\section{Experimental Setup}

For our simulations we have used a source of $10^{18}$ $^{150}$Dy
ions/year, during a total running time of 10 years.

In one case, we use 5 years at an ion acceleration of $\gamma = 195$,
the maximum achievable at CERN's SPS, and 5 years running at $\gamma =
90$, as different as possible from the first one but still avoiding
backgrounds in the detector. In a second case, we combine 5 years of
electron capture at $\gamma = 195$ with a `standard' 5 years of $2.9
\times 10^{18}$ $^6$He ions/year $\beta^-$-beam.

Both scenarios have a 440 kton fiducial mass water Cerenkov detector
located at a distance of 130 km (CERN-Frejus). Information from
appearance and disappearance signals are combined.

\section{Electron Capture and $\beta^\pm$-beam}

The main advantage of an electron capture beam over a $\beta$-beam, for
a similar amount of ion decays, is that all the intensity is peaked at
the energy(ies) of interest. In a $\beta$-beam the broad spectrum implies that
many neutrinos will be produced at energies for which the dependence
with $\delta$ is less pronounced, and/or the cross-section is too low.


\begin{figure}
\begin{tabular}{l}
  \includegraphics[width=12cm]{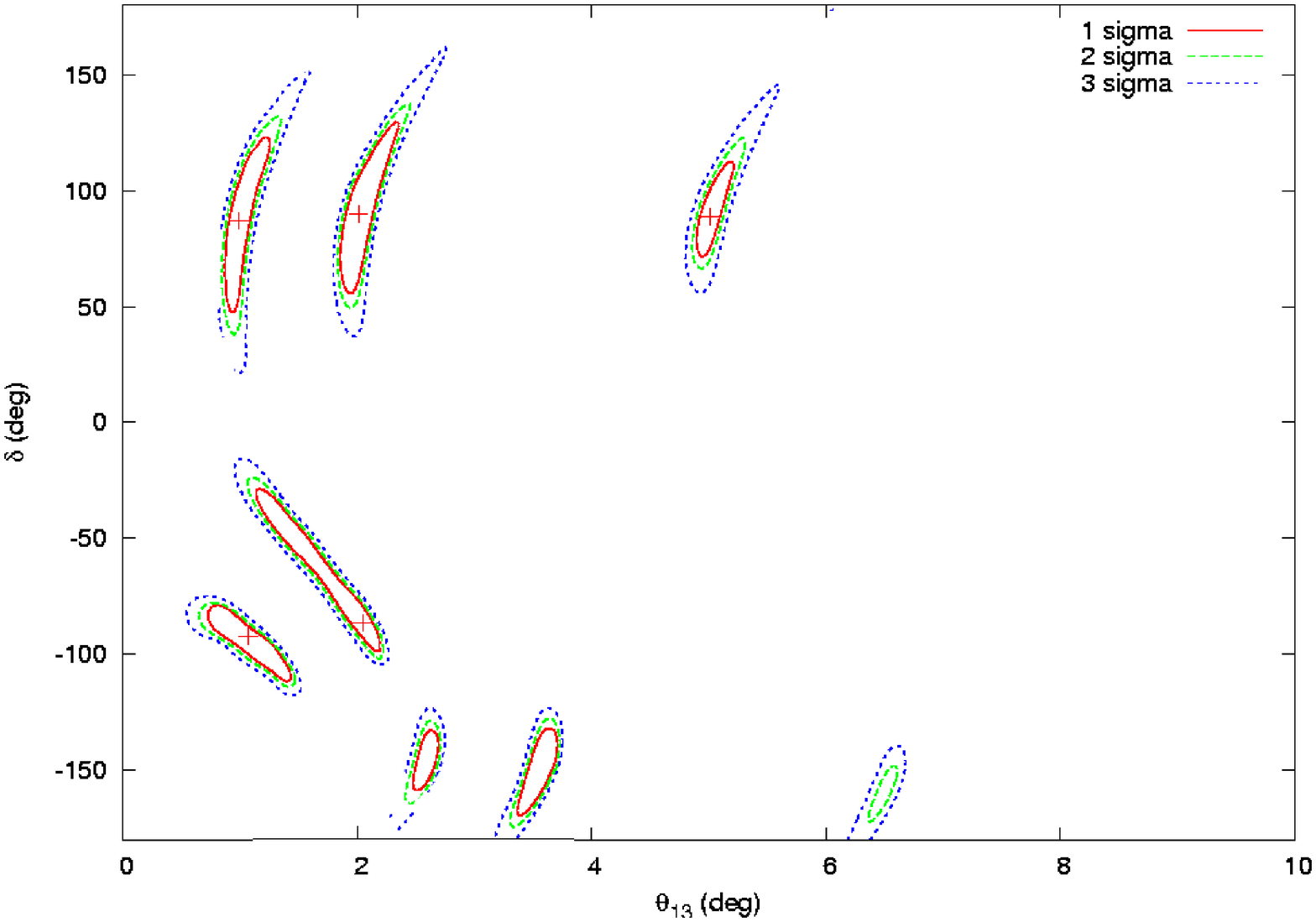} \\
  \includegraphics[width=12cm]{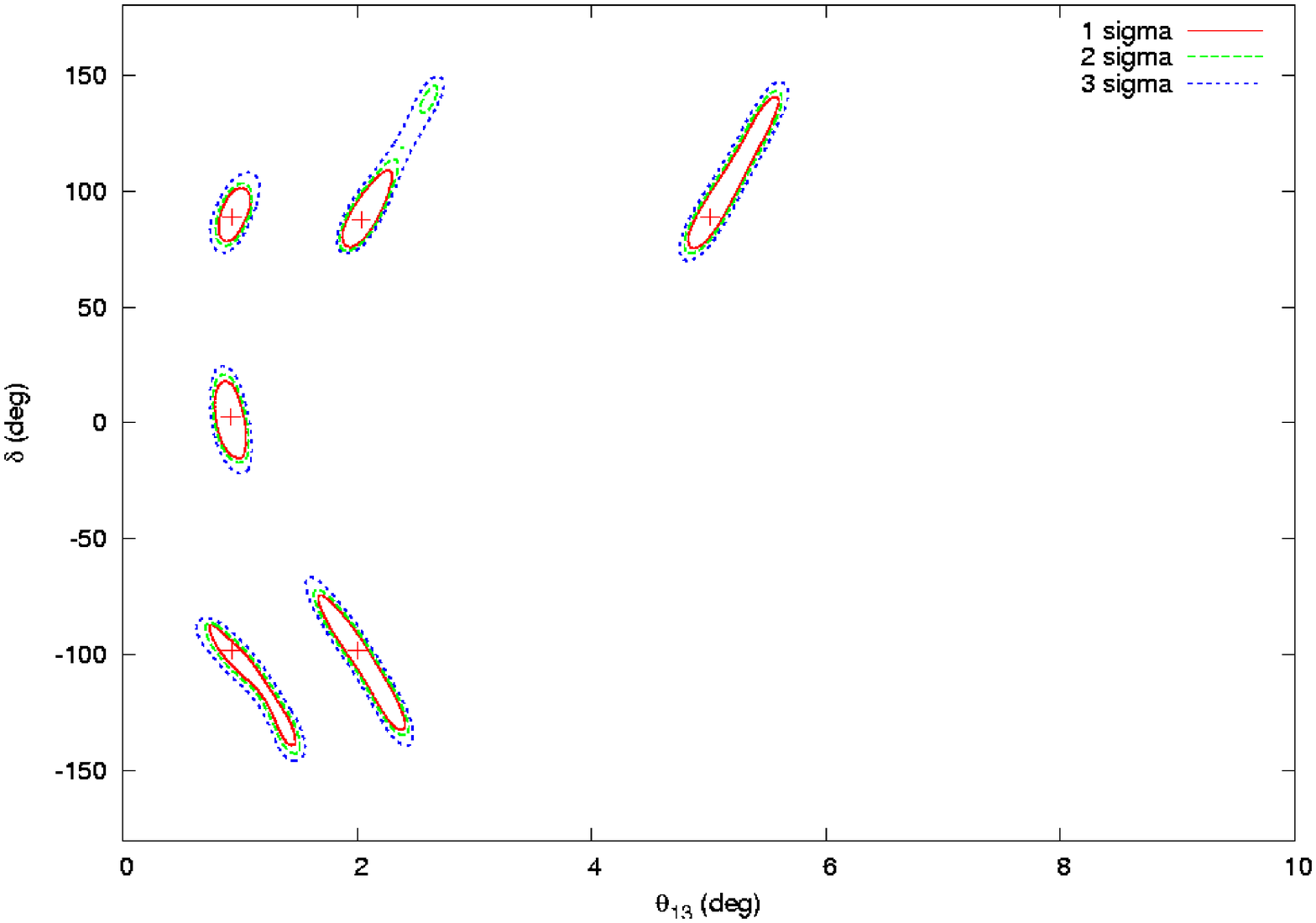} \\
\end{tabular}
  \caption{Combined fits of $\theta_{13}$ and $\delta$ for different
  central values of the parameters. The first (up) is a 5+5 years
  e.c.~at $\gamma = 195, 90$. It has worse resolution in $\delta$ and
  presents degeneracies.  The second (down), 5+5 years $\gamma = 195$ and
  $\beta^-$-beam, show significantly better results.}
  \label{fig:fits}
\end{figure}

On the other hand, it is not possible to produce an antineutrino beam
with this technique. We can still combine runs at different energies
to break the $\delta$ degeneracy, but the shape of the allowed regions
after reconstruction at each energy is slightly similar, and so the
intersection results in fits with sizable uncertainties in
$\delta$. This affects to a (much) lesser amount to the determination
of $\theta_{13}$. However, this also suggests that there must be a
synergy between an electron-capture beam and a $\beta^-$-beam.

\begin{figure}[htb]
  \includegraphics[width=12cm]{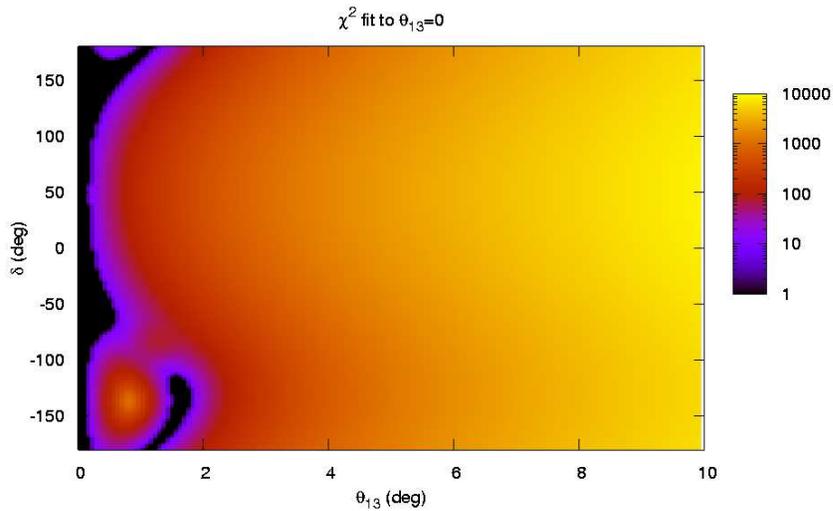}
  \caption{Sensitivity to $\theta_{13} \ne 0$ for 5+5 years of
    e.c.~at $\gamma = 195, 90$. Outside the black region a value of
    $\theta_{13}$ could be distinguished from 0 at $3 \sigma$ C.L.}
  \label{fig:sens_th13}
\end{figure}

Figure (\ref{fig:fits}) shows the results of the fits in both
scenarios for certain chosen center values of the oscillation
parameters. Figure (\ref{fig:sens_th13}) shows the sensitivity to a
$\theta_{13} \ne 0$ achievable with the two electron capture scenario,
which is quite impressive by itself.

Although not presented here, we have also done a preliminary study of
the sensitivity to $\delta$. The scenario with two electron capture
energies give poor results compared with standard proposed
experiments, due to the similar oscillation dependence on $\delta$
even for quite different energies. On the other hand, the combination
with a $\beta^-$-beam shows a very large sensitivity to the discovery
of CP-violation, possibly beyond a Neutrino Factory, an effect mostly
due to the allowed regions of each part intersecting more
perpendicularly for low values of $\theta_{13}$, thus partially
compensating for the simultaneous enlargement of these regions.

\section{Conclusion}

An electron capture neutrino beam is possible thanks to the recent
discovery of heavy ions that decay fast through electron
capture. Preliminary results on the sensitivity of such an experiment
to $\theta_{13}$ and $\delta$ show its potential to compete with a
Neutrino Factory, especially with the second scenario which merges
neutrinos from electron capture with antineutrinos from a
$\beta^-$-beam.

To reach a definite answer on the achievable flux a detailed study of
production cross-sections, target and ion source designs, ion cooling
and accumulation schemes, possible vacuum improvements and stacking
schemes is required. Also, in order to the estimated sensitivity to stand,
we need a careful study which includes the systematics due to the
incomplete knowledge of other oscillation parameters and degeneracies
due to the sign of $\Delta m^2_{23}$ and the quadrant of
$\theta_{23}$.

Nevertheless, our results show the potential interest of the concept
and hopefully encourage the further exploration we feel it deserves.

\end{document}